\documentclass[12pt]{article}
\usepackage{epsfig}

\parskip=\medskipamount	
\renewcommand{\thesection}{\arabic{section}}

\catcode`@=11

\@addtoreset{equation}{section}
\@addtoreset{equation}{subsection}
\def\theequation{\ifnum\value{section}=0 \arabic{equation}\ignorespaces
\else \ifnum\value{section}=-1 A.\arabic{equation}\ignorespaces
\else \ifnum\value{subsection}=0 \thesection.\arabic{equation}\ignorespaces
\else \thesection.\arabic{subsection}.\arabic{equation}\ignorespaces
                             \fi
                        \fi
                   \fi}

{\catcode`\'=\active \def'{{}^\bgroup\prim@s}}

\catcode`@=12

\newcommand{\bq}{\begin{equation}}
\newcommand{\be}{\begin{equation}} 
\newcommand{\fq}{\end{equation}}
\newcommand{\ee}{\end{equation}}
\newcommand{\bqr}{\begin{eqnarray}}
\newcommand{\beqs}{\begin{eqnarray}} 
\newcommand{\fqr}{\end{eqnarray}}
\newcommand{\eeqs}{\end{eqnarray}}

\newcommand{\rf}[1]{(\ref{#1})}

\def\pa{\partial}

\def\bop#1{\setbox0=\hbox{$#1M$}\mkern1.5mu
	\vbox{\hrule height0pt depth.04\ht0
	\hbox{\vrule width.04\ht0 height.9\ht0 \kern.9\ht0
	\vrule width.04\ht0}\hrule height.04\ht0}\mkern1.5mu}
\def\Box{{\mathpalette\bop{}}}                        

\def\Prop{\Delta(m,x)}
\def\Prod{\prod}

\begin{document}
\thispagestyle{empty}

\begin{flushright} 
\begin{tabular}{l} 
UCLA-02-TEP-22\\
hep-th/0209075 \\ 
\end{tabular} 
\end{flushright}  

\vskip .6in 
\begin{center} 

{\Large\bf  Scalar Field Theory in the Derivative Expansion} 

\vskip .6in 

{\bf Gordon Chalmers} 
\\[5mm] 
{\em Department of Physics and Astronomy \\ 
University of California at Los Angeles \\ 
Los Angeles, CA  90025-1547 } \\  

{e-mail: chalmers@physics.ucla.edu}  

\vskip .5in minus .2in 

{\bf Abstract}   
\end{center} 

The quantum correlations of scalar fields are examined as a power series 
in derivatives.  Recursive algebraic equations are derived and determine 
the amplitudes; all loop integrations are performed.  This recursion contains 
the same information as the usual loop expansion.  The approach is pragmatic 
and generalizable to most quantum field theories.  

\setcounter{page}{0}
\newpage 
\setcounter{footnote}{0} 

\section{Introduction} 

Scalar field theories are studied for a large number of reasons.  
However, there are complications due to the complexity of 
multi-loop integrations.  This "bottleneck" prohibits explicit 
results beyond several loop orders.  Better techniques are 
required, and in this work we develop an expansion of loop 
graphs in numbers of derivatives.  This approach has the 
feature that all integrals are performed, and the the computation 
of amplitudes becomes algebraic.  This content has applications 
to high energy physics as well as in condensed matter. Derivative 
expansions have been applied in the context of M-theory 
\cite{Chalmers:2001kx}, N=8 supergravity \cite{Chalmers:2000ks} 
and in scalar field theory \cite{Chalmers:zz} 
(containing some properties of the integrals encountered here).  Prior 
works on simplifing the complexity of diagrammatics is contained in  
\cite{Chalmers:2001cy}, and others referenced therein.

The method is to (re-)write the (loop) expansion in terms of derivatives.  
Consider massive $\phi^n$ theory in $d$ dimensions; adding group 
theory is possible.  It is important to appreciate the fact that 
energy scales are of phenomenological importance; experiments are 
built with greater and greater energies through the course in history.  
>From this point of view, derivative expansions are more natural 
than coupling expansions.  Derivative expansions also commute 
with gauge invariance and the non-perturbative dualities that 
supersymmetric field and string theories possess.  These reasons, 
and the algebraic reduction, motivate the use of derivative 
expansions in the general theory.  

\section{Vertices and Sewing}

The lagrangian is, 

\bqr 
{\cal L} = {1\over 2}\phi\Box\phi + {1\over 2}m^2 \phi^2 + 
\sum {\lambda\over n {!}} \phi^n \Lambda^{4-n} \ , 
\fqr 
and in general we may have derivatives on the vertices if a 
renormalized improved action; we consider dimensions $d\leq 4$.  
In the improved action, all of the irrelevant operators may be 
added.  The standard Feynman rules consist of the propagator 
and vertices.  

The quantum generating functional of the S-matrix is 

\bqr 
\prod_{i=1}^n {\delta\over\delta \phi(x_i)} {\cal L}_{q.c.} \ , 
\fqr  
with, 

\bqr 
{\cal L}_{q.c.} = {1\over 2} \phi(\Box+m^2)\phi + \sum_{d,i_d} 
g^{i_d}_{(d)} {\cal O}^{i_d}_{(d)} \ . 
\fqr  
The operators span all scalar operators of dimension $d$, which 
ranges up to infinity.  At $k^2<\Lambda^2$ (or $m^2$), these are 

\bqr 
\prod_{i=1}^n \bigl[ \left(\Prod_{j=1}^{n_i}
 \pa_{\mu_{\sigma(i,j)}} \right) 
\phi \bigr] {1\over \Lambda^{\tilde d}} \ , 
\fqr 
with $\sum_{i=1}^n n_i + n= {\tilde d}$.  The quantum generating 
functional may be thought of as the infinite number of Feynman 
diagrams expanded at small momenta.  

The general vertex has the form, 

\bqr
t^{\mu_j} \prod_{i=1}^n \left( \prod_{j=1}^{m_i^\pa} 
  \partial_{\mu^{\sigma_{i,j}}} \phi^{m_i^\pa} \right) (m^2)^{-p}  \ , 
\fqr
where the legs are amputated, and $p$ an integer in $4$ dimensions.  
The tensor $t$ contains the coupling dependence from the Lagrangian.  
In a momentum cutoff theory, as 
oppossed to dimensional regularization, the cutoff $\Lambda$ also 
appears on the right hand side of the equation.  If there were 
color in the theory, then the vertex has multiple trace 
structures.  The diagrams seen in the following are planar, but 
are not planar in the color; multi-traces appear in the vertices, 
and are generated from the usual non-planar Feynman diagrams.  

\begin{figure} 
\begin{center} 
\epsfxsize=6cm 
\epsfysize=6cm
\epsfbox[10 0 600 600]{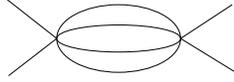}
\end{center} 
\caption{Integrals involved in generating the S-matrix. There are 
an arbitrary number of internal lines, from one to infinity.} 
\end{figure} 

Iterating these vertices gernerate the S-matrix.  Self-consistency 
of the S-matrix through unitarity or through rewriting the usual Feynman 
diagrams generate the diagrams (and integrals) in figure (1), as well 
as the coefficients.  All of the integrals are performed in the 
following, leaving only algebraic recursions for the coefficients.  

The recursion has the form in figure (1) and includes both the sum 
over the intermediate lines and the sum over the derivatives.  
The class of integrals appearing in the sewing are special 
in that they are free-field ones in composite operator calculations; 
these cage diagrams are easily performed in x-space and appear 
complicated in k-space.  For this reason, we work in x-space 
when the integrals are manipulated; the Fourier transform is 
easily implemented.  Another point is that at a given order 
in the external legs, an arbitrary number of internal lines are 
are involved (and arbitrary-point vertex); the lines are correlated 
with the loop order.  

The diagrams in figure (1) evaluate to,  

\bqr 
\sum_L  t^{\mu_\sigma, m_i^\phi} t^{\mu_{\tilde\sigma},{\tilde m}_i^\phi}
\prod_{i=1}^2 \left( \prod_{j=1}^{m_i^\pa} \pa_{\mu_\sigma(i,j)} 
 \phi(k_i)\right)   
  \prod_{i=3}^4 \left( \prod_{j=1}^{{\tilde m}_i^\pa} \pa_{\mu_\sigma(i,j)} 
 \phi(k_i)\right) 
\fqr 
\bqr 
    \times
{1\over L{!}} \langle \prod_{i=1}^L 
  \prod_{j=1}^{m_j^\pa} \pa_{\mu_\sigma(i,j)} \prod_{j=1}^{m_i^\phi} \phi 
  \prod_a^L \prod_{b=1}^{m_b^\pa} \pa_{\mu_{\tilde\sigma(a,b)}}
     \prod_{j=1}^{{\tilde m}_i^\phi} \phi \rangle
+(1\leftrightarrow 3)+(1\leftrightarrow 4) \ , 
\fqr  
with, 

\bqr
\sum m_i^\phi = \sum {\tilde m}_i^\phi = L 
\fqr
The two quantities, the first equation (with coefficient t) and the 
second (product of two t's), are equated; this is the recursion relation. 
It is possible to evaluate all of the integrals in the second equation.

\section{Integrations}

The integrals contain derivatives, and first we simplify these by 
extracting the derivatives.  We simplify the form with the 
identity, 

\bqr 
\partial^\mu \Prop^L = L \left[ (2-d) + m^2 \partial_m^2 \right] 
        \times \left({k^\mu\over k^2}\right) \Prop^L \ .
\label{sub}
\fqr
The massive propagator is 
\bqr
\Prop = (x^2)^{-d/2+1} K_{d/2}(mx)  \ . 
\fqr 
in terms of the modified Bessel function.  
The multiple iterations generate,

\bqr
\int e^{ix\cdot k} \Prod_{j=1}^n \partial_{\mu_j} \Prop^L =
\sum_{\sigma,\sigma'} \prod \eta_{mu_\sigma \nu_\sigma} 
\prod k_{\mu_\sigma'} \left({1\over k^{2n_1+2n_2}}\right) (-2)^{n_2-1}  
\fqr\bqr  
 \times 
  \left[ (2-d) + m^2 \partial_m^2 \right]^n \Prop^L  
\fqr
where the two products contain $n_1$ and $n_2$ terms; we sum over all 
combinations such that $n_1+n_2=n$, found by successively applying 
the substitution rule in eqn \rf{sub}.  Due to the simple topology 
of the diagram only one momenta $k$ labels the external momenta.  This
procedure reduces all tensor integrals to scalar ones, similar to the 
Passarino-Veltman or Feynman-Brown one-loop integral reduction.  

The tensor structure comes from the $t=(k_i+\ldots +k_j)^2$ invariants, 
after transforming the set of internal $k_a$ momenta to x-space, 
$k_a\rightarrow \partial_x$.  We maintain the external lines in
k-space.  

The general integral is, 

\bqr
\sum_L  
  \prod_{i=1}^2 \left( \prod_{j=1}^{m_i^\pa}\pa_{\mu_\sigma(i,j)} 
 \right) \phi(k_i)
  \prod_{i=3}^4 \left( \prod_{j=1}^{{\tilde m}_i^\pa} 
 \pa_{\mu_\sigma(i,j)} \right) \phi(k_i) 
\fqr\bqr 
\times 
\sum_{\sigma,\sigma'} \prod \eta_{mu_\sigma \nu_\sigma} 
\prod (k_1+k_2)_{\mu_\sigma'} (k_1+k_2)^{-2(n_1+n_2)} (-2)^{n_2-1}  
\fqr\bqr  
\times 
\sum_{\tilde\sigma,\tilde\sigma'} 
    \prod \eta_{mu_{\tilde\sigma}\nu_{\tilde\sigma}}
\prod (k_3+k_4)_{\mu_{\tilde\sigma}'} (k_3+k_4)^{-2(n_1+n_2)}(-2)^{n_2-1} 
\fqr\bqr  
\times 
  \left[ (2-d) + m^2 \partial_{m^2} \right]^n 
t^{\mu_\sigma, m_i^\phi} t^{\mu_{\tilde\sigma}, {\tilde m_i^\phi}}
{1\over m_1 {!} \ldots m_i {!}} \quad \int e^{ix\cdot k} \Prop^L  
\fqr 
The integrand reduces to scalar ones, via the contraction of the Pfaffian, 

\bqr 
\langle \prod_i^{m_\phi} \phi(x) \prod_j^{{\tilde m}_\phi} \phi(y)\rangle  
\ . 
\fqr
There is a delta function in $m_i$ and $m_i^\phi$.

The integrals over the products of Bessel functions are evaluated.   
With $k^2=(k_1+k_2)^2=(k_3+k_4)^2$ we have, 

\bqr
\int d^dx \quad e^{ix\cdot k} \Prop^L = (k^2)^{-d/2-L(d/2-1)} 
  \sum_n \left({k^2\over m^2}\right)^n \alpha_n^{(L)} 
\fqr 
\bqr  
= \sum_{n_1\ldots +n_m =n} \prod \beta_{n_i} ~ \left({ L{!}\over 
 n_1{!}\cdots n_m{!}}\right) \int d^dx~ e^{ik\cdot x} x^{nL-(d/2-1)L}  
\label{integral} 
\fqr 
with $\beta$ the expansion coefficients of the Bessel function, and 
\rf{integral} in dimensional regularization, and 

\bqr 
(k^2)^{-d/2-L(d/2-1)} \sum_{m,n} \left({k^2\over \Lambda^2}\right)^m 
\left({k^2\over m^2}\right)^n \alpha_{m,n}^{(L)} \ ,        
\fqr 
in a momentum cutoff scheme.  Both regulators can be applied 
simultaneously.  The integrals follow from a power series expansion of 
the Bessel functions.  Note that the dimension $d=2$ is special 
for two reasons: the integrals are convergent and the factor in the 
substitution rule is nullify, except for the mass differential.    
Exact solution in $d=2$ via recursion is promising.

As the integrals are performed and the recursion is algebraic, now 
it is straightforward to find higher coupling terms from lower coupling.  
The coefficients $t$ have the expansion 

\bqr
t=\sum_{m} \lambda^{m} a_{m} 
\fqr
with the $a_{m}$ generated from the usual Feynman diagrams 
expanded at loop order ${m}$.  It is necessary to resum an 
infinite number of derivatives to rebuild the loop graphs; likewise 
an infinite number of derivative expanded Feynman diagrams are 
required to rebuild the individual terms in the derivative expansion.  
In order to begin the iteration the classical (or quantum improved) 
vertices are required, i.e. couplings 

\bqr
\lambda_3 \phi^3 + \lambda_4 \phi^4 \ , 
\fqr 
and higher-point, or terms with derivatives.  

To show that the derivative expanded graphs reproduce the usual 
coupling ones is straightforward and follows from the Schwinger-Dyson 
or Feynman-Mandelstam tree theorem.  An explicit map to three loops is 
a simple exercise, including color structures.  

\section{Discussion}

The scattering amplitudes of massive $\Phi^n$ theory are examined 
and expressed in the derivative expansion.  Unitarity and equivalence 
with the coupling loop expansion is clear.  All integrals are 
performed, and due to the algebraic nature, the recursive construction 
of the coefficients is well suited to be implemented on a computer; 
integrals, not tensorial algebra, plague numerical calculations.  
The scattering amplitudes have a form similar to a set of matrix theory
calculations.  

These sewing relations are generalizable to gauge theory with
matter, in particular, $N=4$ supersymmetric and quantum chromodynamics, 
and any dynamical system. 

\section*{Acknowledgements} 

GC thanks the DOD, 444025-HL-25619, for support.

\end{document}